\documentclass[a4paper,fullpage]{article}
\usepackage[a4paper, textwidth=15.0cm]{geometry}
\usepackage{graphicx}
\usepackage{amsmath}
\usepackage{bbold}
\usepackage{subcaption}
\numberwithin{equation}{section}
\numberwithin{figure}{section}
\usepackage{authblk}
\usepackage{hyperref}
\usepackage{xcolor}
\usepackage[round, sort]{natbib}
\usepackage{listings}
\usepackage{color}
\usepackage{booktabs}

\usepackage{fancyhdr}
\hypersetup{
    pdftitle={Bayesian Quantile Matching Estimation},
    pdfauthor={Rajbir Singh Nirwan},
    pdfkeywords={Bayes; Machine Learning; Probabilistic Modelling; 
        Finance; Order Statistics; Quantile Matching Estimation},
    linktocpage=true,
    breaklinks=true,
    colorlinks=true,
    linkcolor=blue,
    citecolor=blue,
    urlcolor=blue,
}

\title{Bayesian Quantile Matching Estimation}

\author[,a,b]{Rajbir-Singh Nirwan\footnote{nirwan@fias.uni-frankfurt.de}}
\author[,a,b]{Nils Bertschinger\footnote{bertschinger@fias.uni-frankfurt.de}}

\affil[a]{Goethe University, Frankfurt am Main, Germany}
\affil[b]{Frankfurt Institute for Advanced Studies, Frankfurt am Main, Germany}

\definecolor{codegreen}{rgb}{0,0.6,0}
\definecolor{codegray}{rgb}{0.5,0.5,0.5}
\definecolor{codepurple}{rgb}{0.58,0,0.82}
\definecolor{backcolour}{rgb}{0.99,0.99,0.97}
\lstdefinestyle{custom}{
  language=C++,
  literate={~}{$\sim$}{1},
  backgroundcolor=\color{backcolour},   
  commentstyle=\color{codegreen},
  otherkeywords = {real, vector, matrix, data, model, parameters, transformed,
  functions, generated, quantities},
  keywordstyle=\color{magenta},
  numberstyle=\tiny\color{codegray},
  stringstyle=\color{codepurple},
  emph={normal, cauchy, inv_gamma, bernoulli_logit, gamma,
    lgamma, weibull_lpdf, weibull_rng, weibull_cdf},
  emphstyle=\color{codepurple},	basicstyle={\footnotesize \ttfamily},
  breakatwhitespace=false,         
  breaklines=true,                 
  captionpos=t,                    
  keepspaces=true,                 
  numbers=left,                    
  numbersep=5pt,                  
  showspaces=false,                
  showstringspaces=false,
  showtabs=false,                  
  tabsize=2
}
\lstdefinestyle{oneline}{
  language=C++,
  literate={~}{$\sim$}{1},
  backgroundcolor=\color{backcolour},   
  commentstyle=\color{codegreen},
  otherkeywords = {real, vector, matrix, data, model, parameters, transformed,
  functions, generated, quantities},
  keywordstyle=\color{magenta},
  numberstyle=\tiny\color{codegray},
  stringstyle=\color{codepurple},
  emph={normal, cauchy, inv_gamma, bernoulli_logit, gamma,
    lgamma, weibull_lpdf, weibull_rng, weibull_cdf, lognormal_cdf,
    lognormal_lpdf, lognormal_rng},
  emphstyle=\color{codepurple},	basicstyle={\footnotesize \ttfamily},
  breakatwhitespace=false,         
  breaklines=true,                 
  captionpos=t,                    
  keepspaces=true,                 
  numbersep=5pt,                  
  showspaces=false,                
  showstringspaces=false,
  showtabs=false,                  
  tabsize=2
}

\newcommand*{\Bm}[1]{\boldsymbol{#1}}
\newcommand*{\Bv}[1]{\boldsymbol{#1}}
\newcommand{\X}[1]{X_{(#1)}}
\newcommand{\U}[1]{U_{(#1)}}

\makeatletter         
\renewcommand\maketitle{
{\raggedright
\begin{center}
{\LARGE \bfseries \@title }\\[4ex] 
{\large  \@author}\\[4ex] 
\end{center}}}
\makeatother

\begin{document}

\maketitle
\thispagestyle{plain}

\begin{abstract}
  Due to increased awareness of data protection and corresponding laws
  many data, especially involving sensitive personal information, are
  not publicly accessible. Accordingly, many data collecting agencies
  only release aggregated data, e.g. providing the mean and selected
  quantiles of population distributions.  Yet, research and scientific
  understanding, e.g. for medical diagnostics or policy advice, often
  relies on data access.
  To overcome this tension, we propose a Bayesian method for learning
  from quantile information. Being based on order statistics of finite
  samples our method adequately and correctly reflects the uncertainty
  of empirical quantiles. After outlining the theory, we apply our
  method to simulated as well as real world examples.
  In addition, we provide a python-based package that implements the
  proposed model\footnote{\href{https://github.com/RSNirwan/bqme}
  {https://github.com/RSNirwan/bqme}}.
\end{abstract}

\section{Introduction}

Due to data protection laws sensitive personal data cannot be released
or shared among businesses as well as scientific institutions. While
anonymization techniques are becoming increasingly popular, they often
raise security concerns and have been re-identified in some cases
\cite{Narayanan2010}. To be on the safe side, big data collecting
organisation such as Eurostat (statistical office of the European
Union) or the World Bank only release aggregated summaries of their
data. Instead of individual salary data only selected quantiles
of the population distribution are available.  Thus, for exploratory
analysis as well as statistical modeling, the need for methods which
work on aggregated data is there.  This need can be further expected
to grow as data privacy rules (e.g. GDPR
\citep{EUdataregulations2018}) are becoming more widespread.

Here, we will in particular put emphasis on data where only some
quantile values are available. The matching of quantiles has already
been explored in other contexts.  So far, most of the work in this
area has been built upon a nonlinear regression model where the mean
squared error (MSE) between the cumulative density function (CDF) and
the observed quantile values is minimized. In particular, the Federal
Institute for Risk Assessment in Germany has open sourced an R-package
\citep{rrisk} that fits quantile data using this approach.
Similarly, \citet{Sgouropoulos2015} and \citet{Dominicy2013}
propose to minimize the quadratic distance between quantile statistics
of modeled and actual data. Indeed, \citet{Karian2003} show that
moment matching might be more reliable when using quantiles instead of other
moments.

In this paper, we propose another method for quantile matching 
estimation that is based on the theory of order statistics.
Order statistic is used a lot in the frequentists domain.
\citet{Cohen1998} used order statistics to estimate threshold parameters and
\citet{Chuiv1998} estimated parameters of known distributions. 
\citet{Geisser1998} made use of Bayesian analysis and order statistics for
outlier detection.
Our model also uses order statistics within the Bayesian framework
to estimate parameters by matching observed quantiles.
To the best of our knowledge, this method has not been used in 
this context, but leads to a sound and intuitive noise model.
In particular, we compare this model with 
other noise models, e.g.: the Gaussian noise model, which 
corresponds to the MSE fit mentioned earlier.
We show that our noise model derived from order statistic puts more
emphasis on the tails of the distribution and also captures the
parameter uncertainty better than the Gaussian noise model.

Overall, the goal of this paper is to provide an alternative and principled approach to
quantile matching estimation. For that we utilize the theory of order 
statistics. We start by defining the problem in section \ref{Problem}.
Section \ref{OrderStatistics} introduces the mathematical background needed
to understand our approach. It starts with the uncertainty of the
sample quantiles related to 
the uniform distribution and extends it further to non-uniform
distributions. In section \ref{Experiments} we
conduct experiments to show how our proposed approach performs compared 
to other approaches. 
Finally, section \ref{Conclusion} concludes the paper.

\section{Problem Definition}
\label{Problem}

\begin{figure*}
 \centering
 \includegraphics[width=0.99\textwidth]{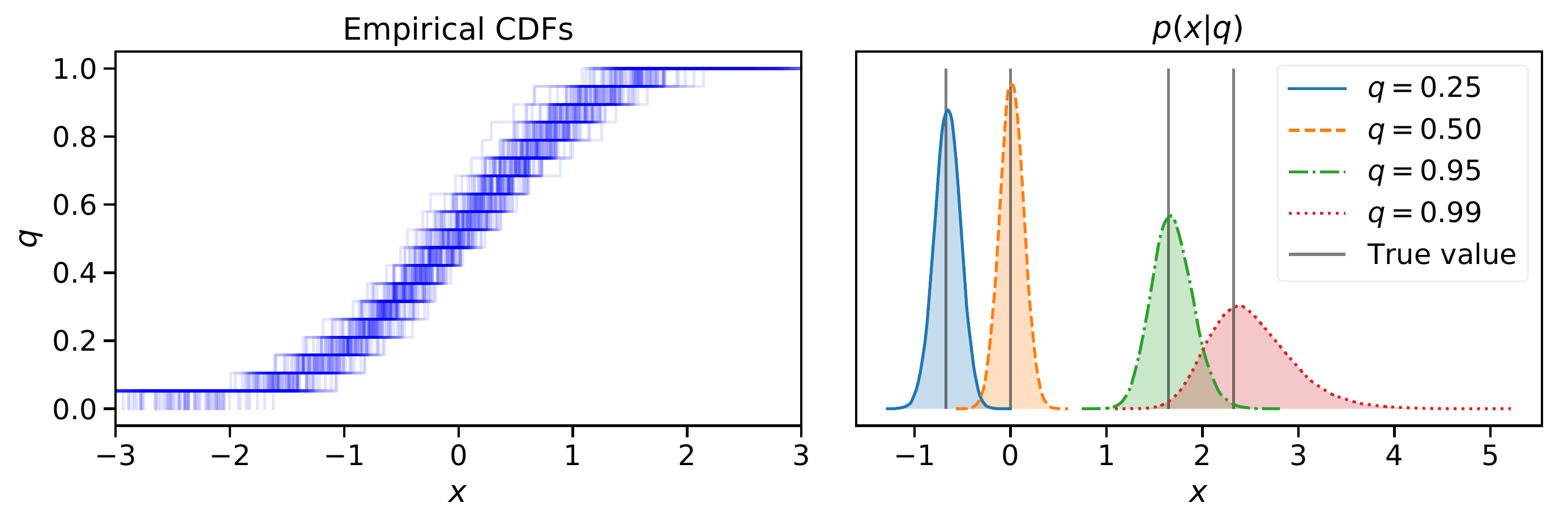}
 \caption{The left panel shows 100 empirical CDFs, each created
        with 20 samples from the standard Gaussian distribution.
    The right panel shows the distribution 
    $p(x|q)$ of the quantile value $x$ for $q = 0.25, 0.50, 0.95$ 
    and $0.99$. This corresponds to the cut of the left panel 
    at the given $q$-values. }
 \label{emp_quan}
\end{figure*}

Suppose we are given $M$ quantiles $\Bv{q} = (q_1, ..., q_M)$,
e.g. for $M=3$ we might be given the $25, 50$ and $75 \%$ quantiles,
and their corresponding empirical values $\Bv{x} = (x_1, ..., x_M)$ in
a sample of $N$ data points drawn from a distribution $p$.
The task is to infer the underlying distribution from this
information\footnote{In particular, we have no access to any of the
  $N$ individual samples, e.g. due to reasons of data
  protection.}. Note that the
quantile values $\Bv{x}$ are calculated from the samples and thus are
not the true values of the underlying distribution.  Other $N$ samples
from the same distribution $p$ will result in different values of
$\Bv{x}$ for fixed quantiles $\Bv{q}$.  That means the quantile values $\Bv{x}$ themselves are
noisy and need to be treated as such. The resulting uncertainty is shown in the left panel
of Figure \ref{emp_quan} showing 100 empirical CDFs of a standard
Gaussian.  For each CDF, we sampled 20 values from a standard
Gaussian, ordered them from low to high, which results in the values of
$\Bv{x}$.  The corresponding $q$-value is just the index $m$ of the
ordered vector $\Bv{x}$ divided by the total number of samples $N=20$,
$q_m = m/N$.  The width (variance) of the CDFs for a fixed quantile
$q_m$ is the uncertainty associated with the corresponding $x_m$.  So,
for a fixed quantile $q$, we have a distribution $p(x|q)$ over $x$,
that we want to model.  This distribution is shown in the right panel
of Figure \ref{emp_quan} for $q = 0.25, 0.50, 0.95$ and $0.99$ for the
standard Gaussian distribution.  Note that around the center ($q=0.5$)
the uncertainty associated with the empirical quantile values $\Bv{x}$
is lower than the uncertainty at the tails ($q=0.95$ and $0.99$). This
difference in the uncertainty of different quantiles $q$ is exactly
what we want our model to capture.

Quantile matching methods as the Gaussian noise model (described in
section \ref{cdfRegressionModel}) do have other prior assumptions for
the uncertainty.  In particular, minimizing the MSE between
theoretical and observed quantiles can be considered as assuming a
fixed Gaussian noise around the fitted CDF (independent of $q$). That,
as we see in the right panel of Figure \ref{emp_quan}, is obviously
not true\footnote{Furthermore, it corresponds to a distribution
  $p(q | x)$, i.e. considering the quantile itself as
  variable. Instead, in most practical applications the quantiles $\Bv{q}$
  are chosen a-priori and their corresponding values $\Bv{x}$ are
  reported. In contrast, our noise model correctly reflects this data
  generating process.}.  This assumption, however, leads to a model
that under emphasizes the uncertainty of the tails and implicitly
downweights the information coming from the samples from the tails. As
we describe in the next section, this problem can be handled using
order statistics.

\section{Order Statistics}
\label{OrderStatistics}
In order to give a solution of the above stated problem, we need to review
some literature on order statistics. In this section we start by looking 
at the order statistics of the uniform distribution and then 
generalize to non-uniform distributions.
At the end of this section
we also introduce the CDF-regression model with Gaussian noise.
In section \ref{Experiments} we will compare the approach for the
regression via order
statistics to the CDF-regression model with Gaussian noise.

\subsection{Order Statistics of a Uniform Distribution}
To fix notation, assume we are given $n$ real-valued iid observations $(X_1, ..., X_n)$ 
from some continuous distribution and the ordered
series of the observations $(\X{1}, ..., \X{n})$, where
$\X{1} \le \dots \le \X{n}$. The $k$-th order statistic
of that sample is equal to its $k$-th smallest value $\X{k}$.
As stated in the problem definition (section \ref{Problem})
we are interested in the distribution of the order
statistic $p(\X{k}|k)$\footnote{
Note that the maximum, minimum, median and other quantiles are also
order statistics, since $\X{1} = \min(X_1, \dots, X_n)$, 
$\X{n} = \max(X_1, \dots, X_n)$, $\X{(n+1)/2} = 
\text{median}(X_1, \dots, X_n)$ for odd $n$. There is an ambiguity
for even $n$. But that is not important for us right now.}.

To get an intuition of the order statistics we will start with 
the uniform distribution on the interval $[0,1]$. 
Thus, let $(X_1, \dots, X_n)$ be iid samples from a $\mathrm{Uniform}(0,1)$ distribution.
We want to find the PDF $p(\X{k})$ and the CDF $P(\X{k})$ 
of the $k$-th order statistics $\X{k}$. For $x \in [0,1]$, the
CDF $P(\X{k}) = P(\X{k} < x)$ has the following interpretation:
if the event $\X{k} < x$ occurs, then there are at
least $k$ many $X_i$'s that are smaller or equal to $x$. 
Thus, drawing all the samples $(X_1, \dots, X_n)$ can be
seen as Bernoulli trials, where success is defined
as $X_i < x \ \forall \ i \in \{1, \dots, n\}$.
$P(\X{k} < x)$ is then defined as at least $k$ successes and
has the following form
\begin{equation}
    P(\X{k} < x) = \sum_{i = k}^n {n \choose i} x^i (1-x)^{n-i} \; .
    \label{cdfOSU}
\end{equation}

The PDF is given by the derivative of equation (\ref{cdfOSU}) and takes
the form
\begin{align}
    p(\X{k} = x) = n {n-1 \choose k-1} x^{k-1}(1-x)^{n-k} \; ,
    \label{pdfOSU}
\end{align}
which is the Beta distribution $\text{Beta}(k, n-k+1)$. 
It has also a convenient interpretation: All $n$ samples that
can result in the $k$-th order statistic, one of them becomes 
$\X{k}$ and out of the $n-1$ left,
there must be exactly $k-1$ successes (defined as $X_i < \X{k}$).
So, for samples from a 
standard Uniform distribution the $k$-th order statistic $\X{k}$ is
Beta-distributed  $\X{k} \sim \text{Beta}(k, n-k+1)$.

\subsection{Generalization of Uniform Order Statistics}
In the more general case, where $\X{i}$s are random
samples drawn from a continuous non-uniform distribution with density $f_x$,
we again ask for the $k$-th order statistics $p_x(\X{k})$.
In this case, we just transform the samples such that they become uniformly distributed
samples and then use the procedure described in the previous section.
However, since we are transforming a random variable, we have to correct
for the resulting change of measure by multiplying with the absolute Jacobian.

Given the cumulative distribution function $F_x$, we obtain
$U_i = F_x(X_i)$. $U_1, ..., U_n$ correspond to samples from
the standard uniform distribution. Again, $\U{k}$ will
be Beta distributed
\begin{equation}
    \U{k} = F_x(\X{k}) \sim \ \text{Beta}\left(F_x(\X{k})|k, n-k+1 \right) := p_u(\U{k}).
 \label{xToUniform}
\end{equation}
Given a probability $p_u(u)$ on $u$ and an invertible map $g$, so that
$u = g(x)$, the density $p_x(x)$ on $x$ is given by
\begin{equation}
p_x(x) = p_u\left(g(x)\right) \left| \frac{dg(x)}{dx} \right| \; .
\end{equation}
In our case the invertible map is the cumulative  distribution function $F$.
So, we get
\begin{equation}
    p_x(\X{k}) = p_u\left(F_x(\X{k})\right) 
    \left| \frac{\text{d}F_x(x)}{\text{d}x} \right|_{x=\X{k}} \; .
\end{equation}
Using eq. (\ref{xToUniform}) and since $|\text{d}F_x(x)/\text{d}x| = f_x(x)$,
we obtain
\begin{equation}
    p_x(\X{k}) = \text{Beta}\left(F_x(\X{k})|k, n-k+1 \right)
    f_x(\X{k}) \; ,
 \label{generalLikelihood}
\end{equation}
which is the $k$-th order statistic of any random variable $X$ with
the PDF $f_x$ and CDF $F_x$.

\subsection{Joint Distribution of the Order Statistics}

Even if the underlying samples are independent, their order statistics will not be. Thus,
in order to learn from observed (quantile values) $\Bv{x} \in \mathbb{R}^M$ and
their corresponding order $\Bv{k} \in \mathbb{N}^M$
we need to know the joint PDF of the order statistics
$p_x(\X{1}, \ldots, \X{M}|\theta) \neq \prod_k p_x(\X{k} | \theta)$.
In particular, if $i < j$, $\X{i}$ will for sure be smaller than $\X{j}$.
We start with the joint distribution of two such observations.

\begin{figure*}
    \centering
    \includegraphics[width=0.90\textwidth]{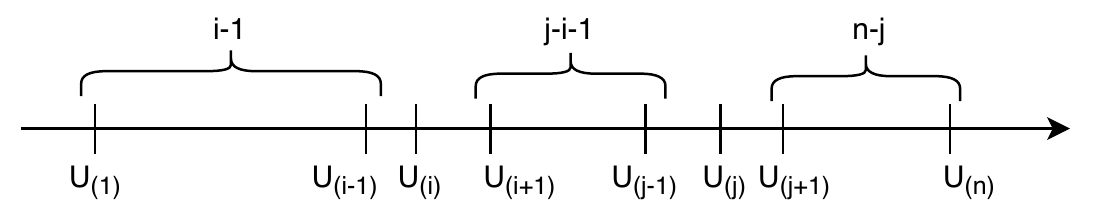}
    \caption{Illustration of the 
    PDF of the joint order statistics (equation \ref{eq::2jointos}).}
    \label{fig::jointOS}
\end{figure*}

The application of the
CDF $F$ to all of the observed quantile values $\Bv{x}$,
leads to a uniformly distributed random vector $\Bv{U} = (U_1, \dots, U_M)$.
The joint PDF of two order statistics $\U{i}$ and $\U{j}$, where
$\U{i} < \U{j}$ then takes
the following form
\begin{equation}
 p_{\U{i}, \U{j}}(u, v) = n! \frac{u^{i-1}}{(i-1)!} \frac{(v-u)^{j-i-1}}{(j-i-1)!}
    \frac{(1-v)^{n-j}}{(n-j)!} \; ,
    \label{eq::2jointos}
\end{equation}
where $u$ and $v$ correspond to the observed values of $\U{i}$ and $\U{j}$,
the first fraction ${u^{i-1}}/{(i-1)!}$ is proportional to the binomial
distribution of $i-1$ of the samples being smaller than $u$. 
The second term ${(v-u)^{j-i-1}}/{(j-i-1)!}$ corresponds to 
$j-i-1$ of the samples being in the interval $(u, v)$ and 
the last term ${(1-v)^{n-j}}/{(n-j)!}$ corresponds to $(n-j)$ samples
being greater than $v$. This process is illustrated in Figure \ref{fig::jointOS}.

The procedure can be extended to $M$ observed values from a uniform
distribution $\Bv{u} = (u_1, \dots, u_M) \in \mathbb{R}^M$ with their order 
$\Bv{k} = (k_1, \dots, k_M) \in \mathbb{N}^M$ 
and leads to the
following joint PDF
\begin{equation}
 p_{\Bv{U}}(\Bv{u}|\Bv{k}) = c u_1^{k_1-1} (1-u_M)^{n-k_M}
    \prod_{m=2}^M (u_m - u_{m-1})^{k_m-k_{m-1}-1},
    \label{jointUniformOS}
\end{equation}
where the normalization constant $c$ is given by (see Appendix 
\ref{ANormalizationConstatnt})
\begin{equation}
    c = \frac{n!}{(k_1-1)!(n-k_m)! \prod_{m=2}^M(k_m-k_{m-1}-1)!} \; .
    \label{normalizationConstant}
\end{equation}

From here, it is easy to extend equation (\ref{jointUniformOS}) to
samples from a non-uniform distribution with PDF $f$ and CDF $F$.
As already mentioned earlier, applying $F$ to the samples will
convert them to the desired samples from a uniform distribution
and we can use equation (\ref{jointUniformOS}) adjusted with
the Jacobian correction. So, for observations
$\Bv{x} \in \mathbb{R}^M$ from a PDF $f$, their corresponding order 
$\Bv{k} \in \mathbb{N}^M$ and the 
total number of observations $n$, we get the following 
joint order statistics
\begin{equation}
 p_{\Bv{X}}(\Bv{x}|\Bv{k}) = c F(x_1)^{k_1-1} (1-F(x_M))^{n-k_M}
    \prod_{m=2}^M (F(x_m) - F(x_{m-1}))^{k_m-k_{m-1}-1}
    \prod_{m=1}^M f(x_m).
    \label{jointFOS}
\end{equation}

If instead of the order $\Bv{k}$ we are given the corresponding
quantile information $\Bm{q} = (q_1, \dots, q_M) \in [0, 1]^M$,
we just need to replace the $k_m$ with $nq_m$ 
in equation (\ref{jointFOS}).

\subsection{Fit of Non-Uniform Distributions given Quantile Information}

By utilizing the order statistics of a random variable, we can now 
use the observed quantiles and infer the underlying distribution.
Given $M$ observed quantiles, which are based on $N$ samples, we 
denote as $\Bv{q} = (q_1, q_2, \dots, q_M) \in [0, 1]^M$ the 
quantiles and $\Bv{x} = (x_1, x_2, \dots, x_M) \in \mathbb{R}^M$ 
their corresponding empirical values. $k$ as defined previously, is simply the product of 
a quantile $q_i$ and the number of total samples $N$,
i.e.: $k_i = q_i N$\footnote{Note $k_i$ is a positive integer 
but $q_i$ can be any number between 0 and 1. Thus, by not setting
$k_i$ as $\lfloor q_i N \rfloor$ or $\lceil q_i N \rceil$ but as
$k_i = q_i N$ we have an interpolation.}.

With these definitions, the model likelihood becomes
\begin{align}
    p_{\Bv{X}}(\Bv{x}|\Bv{q}, \Bv{\theta}, N) = 
&c F_{\Bv{\theta}}(x_1)^{q_1N-1} (1-F_{\Bv{\theta}}(x_M))^{N-q_MN} \nonumber \\
&\prod_{m=2}^M (F_{\Bv{\theta}}(x_m) - F_{\Bv{\theta}}(x_{m-1}))^{q_mN-q_{m-1}N-1}
\nonumber \\
&\prod_{m=1}^M f_{\Bv{\theta}}(x_m) \; ,
 \label{os_ll}
\end{align}
where $F_{\Bv{\theta}}$ is the CDF parameterized by $\Bv{\theta}$.
From here on, one can either maximize equation (\ref{os_ll}) 
with respect to the parameter $\Bv{\theta}$, or infer the full
posterior $p(\Bv{\theta}|\Bv{x}, \Bv{q}, N)$.
We choose the latter, since it provides the full distribution 
of all parameters and thus also includes their estimation uncertainty. 
For the fully Bayesian treatment of the model, we 
assign a prior distribution $p(\Bv{\theta})$ over the parameters $\Bv{\theta}$ 
and infer the posterior.
The posterior is then given by Bayes rule
\begin{equation}
 p(\Bv{\theta}|\Bv{x}, \Bv{q}, N) = 
 \frac{p(\Bv{x}|\Bv{\theta},\Bv{q},N) p(\Bv{\theta})}{p(\Bv{x})}\; ,
\end{equation}
which is intractable. There are several methods to approximate the 
posterior e.g.: Markov Chain Monte Carlo (MCMC), Variational Bayes 
(VB), etc..
We implement the model in the probabilistic programming 
language Stan \citep{stan}, where both, MCMC (NUTS version of 
HMC) and VB with Gaussian approximation to the posterior, can be 
used out of the box. For our experiments we use MCMC.

\subsubsection{Generative Model}
Given $N$, $\Bv{q}$, $\Bv{x}$ and the probability function $f_{\Bv{\theta}}$,
parameterized by $\Bv{\theta}$, we obtain the following generative model:
\begin{align}
  \Bv{\theta} &\sim p(\Bv{\theta})  \nonumber \\
  \qquad \Bv{x} &\sim p_{\Bv{X}}(\Bv{x}|\Bv{q}, \Bv{\theta}, N) \; ,
\end{align}
where $F_{\Bv{\theta}}$ is the CDF of $f_{\Bv{\theta}}$.
$p(\Bv{\theta})$ is the prior distribution for our parameters and
$p_{\Bv{X}}(\Bv{x}|\Bv{q}, \Bv{\theta}, N)$ is the likelihood given
in equation (\ref{os_ll}). For all the experiments in 
section \ref{Experiments} we took a very broad Gaussian
prior for all parameters (see below).
The Stan code for the model is provided in the supplementary material.
In addition, we have written a BQME-package in python
(\href{https://github.com/RSNirwan/bqme}
{https://github.com/RSNirwan/bqme}).
The package allows for an easy specification of the model
(prior and likelihood) and inference in just a few lines of code.
After inference, it also allows the user to generate new samples and evaluate
the density of unseen data.

For all presented examples and most practical purposes, i.e. where $M$
is small, the code runs almost instantaneously. In particular, by equation
(\ref{os_ll}), computing the likelihood has linear complexity requiring $O(M)$ CDF and PDF
evaluations.

\subsection{CDF Regression Model}
\label{cdfRegressionModel}

Another approach for quantile-matching estimation (QME) which is
widely used nowadays, is based on fitting the CDF to observed quantiles via
mean squared error (MSE) minimisation. Given the quantiles $\Bv{q}$ and the values at the quantiles 
$\Bv{x}$, the idea is to choose a parametric form of the distribution
$f_{\Bv{\theta}}$ (parameterized by $\Bv{\theta}$) and find the parameters such that the MSE
\begin{equation}
  \label{eq:msefit}
  \min_{\Bv{\theta}} \sum_{m = 1}^M (q_m - F_{\Bv{\theta}}(x_m))^2
\end{equation}
is minimized. Thus, one basically fits the CDF $F_{\Bv{\theta}}$
of $f_{\Bv{\theta}}$ to the observed data ($\Bv{q}$, $\Bv{x}$).  Note
that the MSE error function corresponds to the log-likelihood of a
Gaussian noise model. Thus, we can equivalently consider the maximum
likelihood estimate (MLE) for the parameters $\Bv{\theta}$
for a model with likelihood
\begin{equation}
 p(\Bv{q}|\Bv{\theta}, \Bv{x}, \sigma_{\text{noise}}^2) = \prod_{m=1}^M
 \mathcal{N}
 \left( \Bv{q}_m|F_{\Bv{\theta}}(\Bv{x}_m), \sigma_\text{noise}^2 \right) \; .
 \label{cdffit_ll}
\end{equation}

By comparing equation (\ref{cdffit_ll}) to (\ref{os_ll}), we see that
the modelling of the uncertainty is quite different.  Whereas the
regression of quantiles puts the noise on the quantiles $\Bv{q}$
(equation \ref{cdffit_ll}), the solution with the order statistics
treats the observed quantile values $\Bv{x}$ as noisy (equation
\ref{os_ll}).  On the one hand, this reverses the data generating
process where in practice the quantiles $\Bv{q}$ are chosen a-priori
and their corresponding values $\Bv{x}$ are reported.  On the other hand,
it leads to a different penalty for the deviation of the regression
function. As already stated in section \ref{Problem}, this model
assumes a fixed Gaussian noise around the fitted CDF, which is
independent of $q$ or $x$. This assumption, however, leads to an
underestimation of the noise uncertainty, especially at the tails and
leads to another fit than the order statistics.
We further discuss these points in the next section.

\section{Experiments}
\label{Experiments}

In this section we conduct experiments on quantile matching estimations
and discuss the results. We will start by looking at the 
posterior distribution of the parameters of the PDF for both noise
models (order statistics and the Gaussian noise model). 
Subsequently, we will discuss their sensitivity to the change of single 
data points, which is also related to the robustness of the models.
At the end, we analyse how the 2 noise models emphasise
different regimes (tails or the center of the PDF) of the observed data.
The prior on the parameters of both models in all experiments is 
set to a very broad Gaussian with mean 0 and standard deviation 100.
The code for the simulations is available on Github:
\href{https://github.com/RSNirwan/BQME\_experiments}
{https://github.com/RSNirwan/BQME\_experiments}.

\subsection{Bayesian Quantile Matching Estimation}
In this subsection we will deal with the case of correctly specified models, where 
the data generating distribution is from the same class as the 
distribution that we use to fit the data. In real world applications
one does not know the data generating distribution and the model
distribution may differ from it. In section \ref{sec::penalty}
we consider the case of misspecified models and discuss how this
situation can be detected.

\subsubsection{Gaussian Fit}
We will start by fitting a Gaussian distribution to the 
observed data. The data are generated by taking $N$ samples from a 
Gaussian with known parameters $\mu$ and $\sigma$. Afterwards we 
take $M$ quantile values $\Bv{x} \in \mathbb{R}^M$ for given 
quantiles $\Bv{q} \in \mathbb{R}^M$. The tuple $(\Bv{x}, \Bv{q}, N)$
is the ``observed data'' that we fit by a Gaussian PDF. 
This is done with order statistics (equation \ref{os_ll})
as well as with the Gaussian noise model (equation \ref{cdffit_ll}).

\begin{figure}
 \centering
 \includegraphics[width=0.7\textwidth]{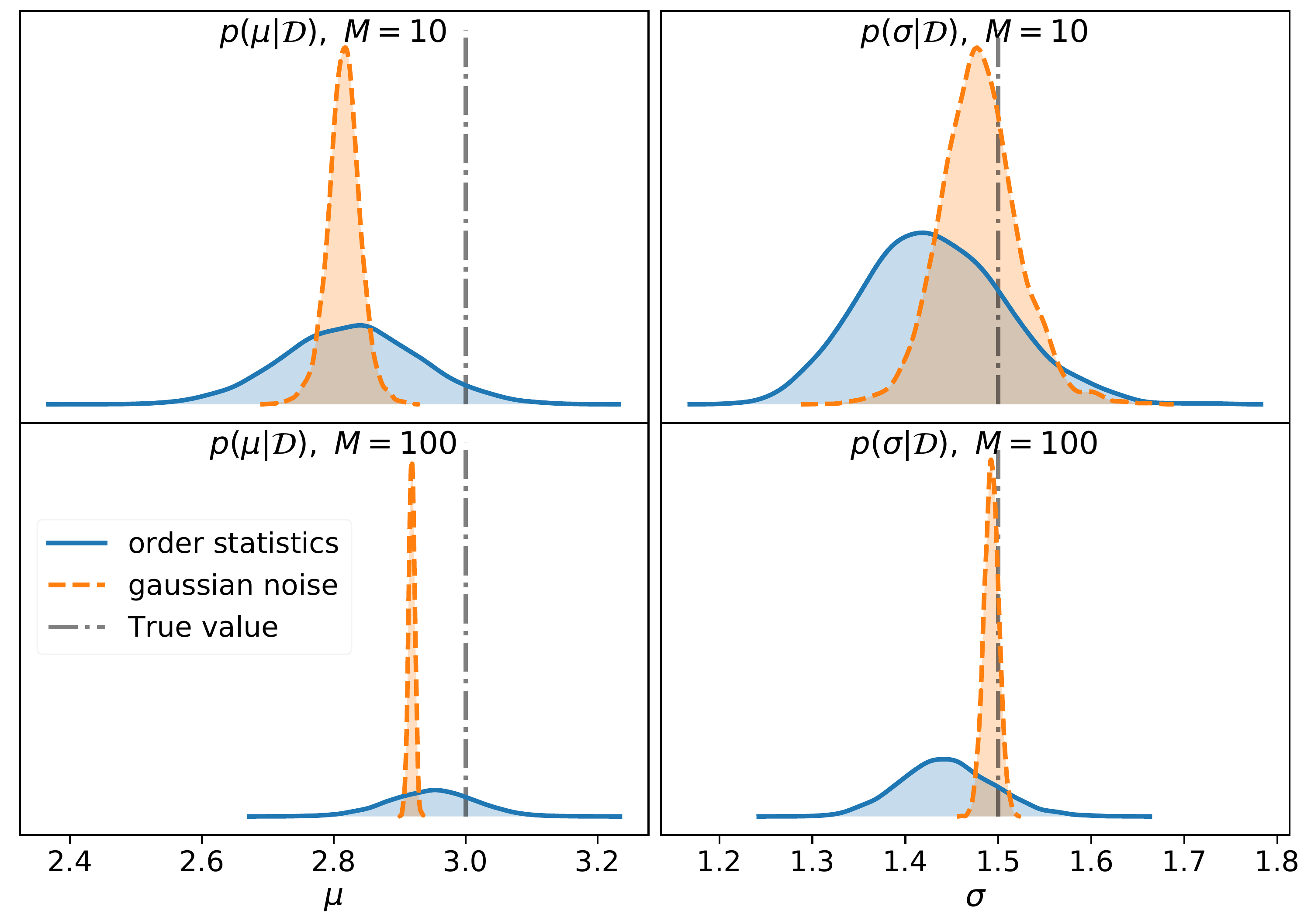}
 \caption{The posterior parameter distribution for a fit with a 
    Gaussian distribution. Top row shows the results for $\mu$ and $\sigma$
    for $M=10$ and the bottom row for $M=100$. $\Bv{q}$s were chosen 
    equidistantly between 5\% to 95\%, $N=200$ for $M=10$ and
    $N=500$ for $M=100$ and
    the true values for $\mu$ and $\sigma$ are 3.0 and 1.5, respectively.}
 \label{bqme_gaussian}
\end{figure}

Figure \ref{bqme_gaussian} shows the sampled posterior for $\mu$ and $\sigma$. 
The kernel density estimate (KDE) of the samples is plotted 
and true values of the 
data generating distribution (which is also a Gaussian) are shown
with vertical black line and were set to $\mu=3.0$ and $\sigma=1.5$.
The top row shows the results for $N=200$ and $M=10$ and the 
bottom row shows the results for $N=500$ and $M=100$. $\Bv{q}$ is chosen
equidistantly between 5\% and 95\% for both plots.
As one can see, the Gaussian noise model clearly
underestimates the uncertainty for $\mu$ compared to the order statistic 
model. The probability mass the Gaussian noise model puts on the 
true value is almost zero, which is not a desirable property 
for any model.

\subsubsection{Dependency on $N$}
When we fit a distribution to some observed quantile values, we 
would ideally expect an effect of the number of samples the observed quantile values 
are based on. The higher the
sample size, the closer (less uncertain) 
observed sample quantile values will be to the true quantile values of the
underlying distribution. This is taken care of in the order statistic.
The likelihood of the order statistics (equation \ref{os_ll}) is 
dependent on $N$. The higher $N$, the more accurate the
parameters of the fitted distribution will be. In this subsection we 
will look at the change of the posterior distribution as a function of 
$N$. We fix $M$ and $\Bv{q}$. $\Bv{x}$ is set to the true quantile 
values of a Gaussian. The only parameter we change is $N$.

\begin{figure}
 \centering
 \includegraphics[width=0.8\textwidth]{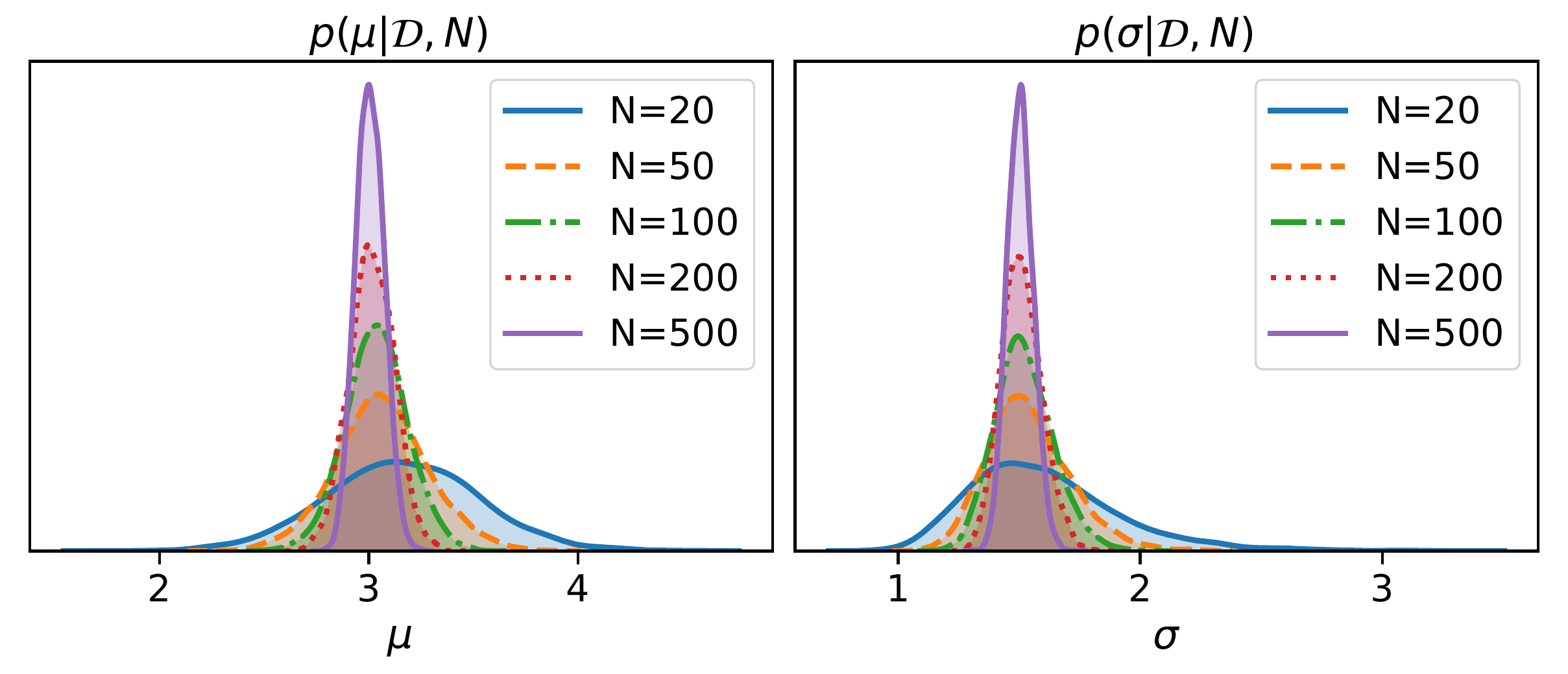}
 \caption{Posterior parameter distribution for a Gaussian
    fit for different $N$. With higher $N$ the uncertainty 
    in the posterior is reduced significantly.}
 \label{dependencyN}
\end{figure}

Figure \ref{dependencyN} shows the results for $\mu=3.0$ and
$\sigma=1.5$, $M$ was set to 10 and the quantiles $\Bv{q}$ were chosen
equidistantly between 5\% and 95\%. As expected, with higher $N$ the
uncertainty in the posterior is decreasing.  This is driven by the
model likelihood in equation (\ref{os_ll}) which exactly and correctly
captures the dependency on $N$ in the uncertainty of the empirical
quantiles.  In contrast, the Gaussian noise model would only
indirectly respond to increasing sample sizes as the empirical CDF
more closely tracks the theoretical one and the MSE is reduced,
i.e. $\sigma_{\mathrm{ML}}^2 \to 0$ for $N \to \infty$.

\subsection{Penalty for OS and CDF-fit}
\label{sec::penalty}
As already mentioned, the penalty of the observed data points
$(\Bv{q}, \Bv{x}, N)$ to the regression function is quite 
different by choosing either the order statistics error probability
(equation \ref{os_ll}) or the Gaussian error probability 
(equation \ref{cdffit_ll}).
This can be illustrated by looking at the true and the 
matched quantile plots.

We take $N$ samples from the data generating distribution,
calculate $M$ quantile values $\Bv{x}$ for given quantiles $\Bv{q}$.
Subsequently, we fit the desired distribution to
$(\Bv{q}, \Bv{x}, N)$ via the order statistics 
and the Gaussian error function.
Ideally, one would take the same distribution to fit the data as the
data generating distribution, but this cannot be 
guaranteed, since for most real world cases one rarely 
knows its form.

\begin{figure}
 \centering
 \includegraphics[width=0.8\textwidth]{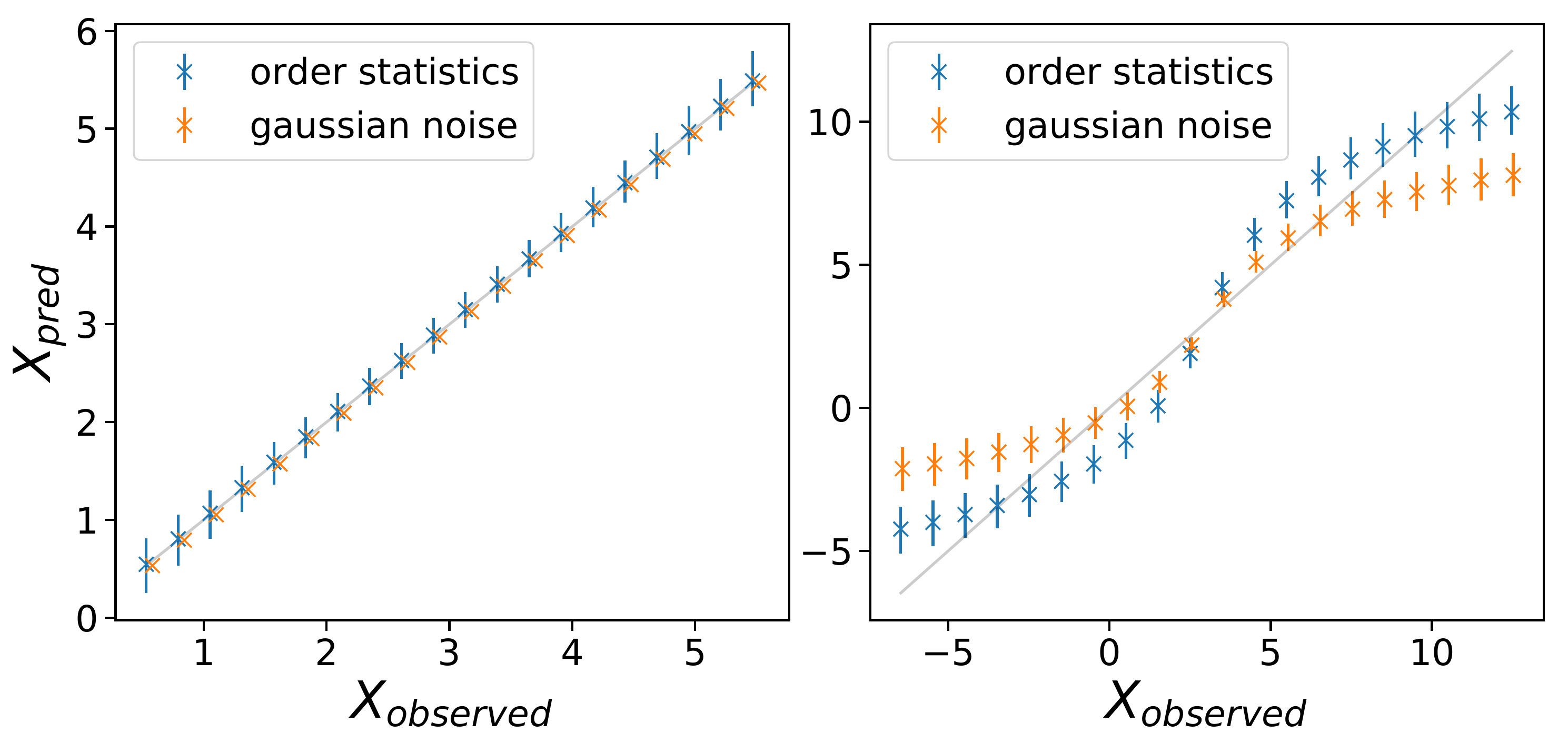}
 \caption{Correctly specified (left) and misspecified model (right) fitted with 
    the order statistics and Gaussian noise. In the specified case,
    both models perform quite well, but the Gaussian noise underestimates
    the uncertainty. In the misspecified case, both models put emphasis on
    different parts of the data.}
 \label{xt_xpred_misspecified}
\end{figure}

Figure \ref{xt_xpred_misspecified} shows the results for a fit with
the Gaussian probability function.
Left hand side shows the result for a correctly specified model, where the 
fitted distribution is from the same class as the data generating distribution.
Here, both are Gaussian and the location and scale parameters
for the data generating Gaussian distribution were set to 3.0 and 1.5.
The right hand side shows the results for a misspecified model,
where the
data generating distribution is not from the same class as the 
distribution that we fit to the data. Here, the data generating 
distribution was a Cauchy with location and scale parameters set to
3.0 and 1.5 and we fitted a Gaussian to the 
observed quantile values.
In each case, we took $N=200$ and 20 equidistant quantile values in the range from 
5\% to 95\%. Note that since the Cauchy distribution is heavy tailed,
the scale of the coordinate axes is not the same in both figures.

Both models perform quite well in the case of the specified model,
but the order statistics clearly capture the uncertainty better
than the Gaussian noise model, which underestimates it.
In the misspecified case, each
of the model focuses on different parts of the data.
While the Gaussian noise model emphasises the central part
of the distribution, 
the tails are better captured by the order statistics. 

\begin{figure*}
 \centering
 \includegraphics[width=0.99\textwidth]{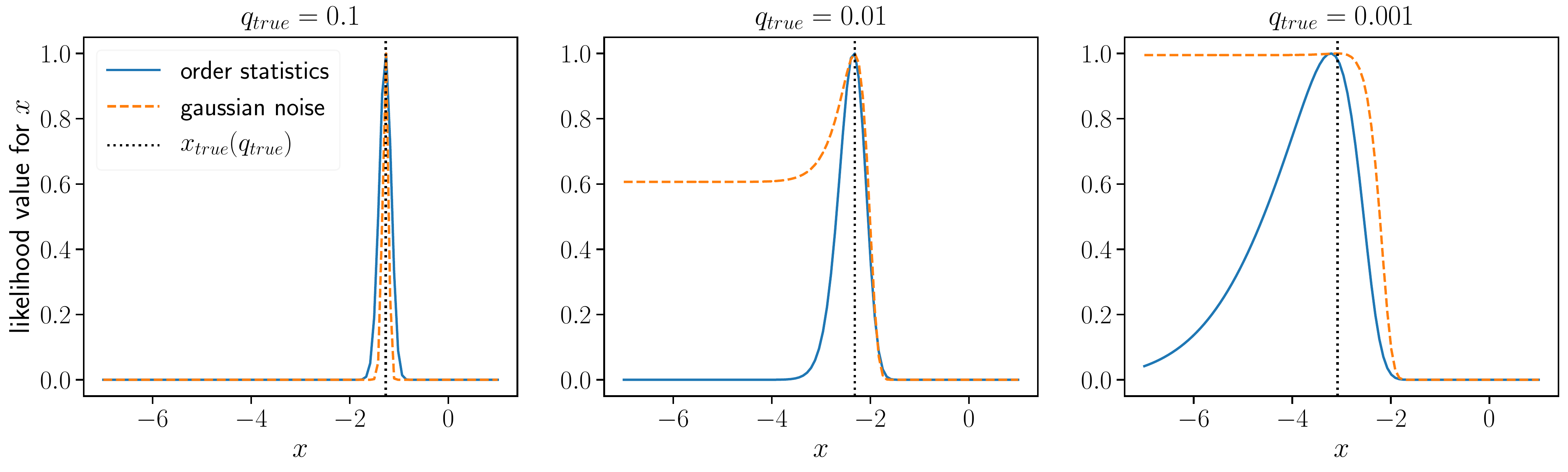}
 \caption{ Likelihood of Gaussian noise model and order statistics as
   a function of $x$ at different fixed quantiles $q$. Note that the
   likelihood of the Gaussian noise model does not decrease to zero
   even for arbitrarily large deviations from the true value
   $x_{true}$ in the (left) tail.
 }
 \label{differentnoises}
\end{figure*}

Note, that we are fitting data from a Cauchy distribution with 
a Gaussian distribution. Because of the very different tail
behavior (heavy tails of Cauchy and light tails of the Gaussian)
those will never be fitted correctly. The Gaussian noise model, which 
corresponds to the MSE function, does not put much emphasis
on the tails. In the tails the difference between even the CDF of a very 
heavy tailed distribution (as the Cauchy) and a very light tailed distribution (as the
Gaussian) is negligible compared to their differences around the mode.
Thus, the contribution of the tails to the overall MSE
is very small and leads to the neglect of the tails.
In contrast, the order statistics correctly weights the information
from all observations and strives to increase the variance of the Gaussian to 
capture some of the heavy tail behavior.
This is shown in 
Figure \ref{differentnoises}. We plotted the likelihood of the
order statistics (blue) and the Gaussian noise model (orange)
for a standard Gaussian $\mathcal{N}(0, 1)$ with PDF $f(x)$ and CDF $F(x)$. 
In particular, we
plotted the likelihoods (normalized such that the maximum is at 1)
as a function of $x$
\begin{align}
    p_{\text{os}}(x|q) &\propto F(x)^{qN-1} 
    (1 - F(x))^{N-qN} f(x) \label{eq::os} \\ 
    p_{\text{gn}}(q|x) &\propto \exp \left\{
    - \frac{1}{2 \sigma_{\text{noise}}^2} (F(x) - q)^2\right\} \label{eq::gn}
\end{align}
for three fixed quantiles $q\in\{0.1, 0.01, 0.001\}$.
$p_{\text{os}}$ is the likelihood for the order statistics and
$p_{\text{gn}}$ is the likelihood for the Gaussian noise model.
The black vertical line shows the quantile value $x_{\text{true}}$ 
for the fixed quantile $q_{\text{true}}$ and the blue and orange 
lines show the score that an estimated value $x$ corresponding to a $q$ 
would get.
The order statistics show a peak at the true value 
$x_{\text{true}}$ and decreases on both sides even if
$x_{\text{true}}$ is set
further in the tails $(q = 0.1, 0.01, 0.001)$. The Gaussian noise model 
on the other hand does not decrease to zero even far away from
the true quantile value (middle and right plot). 
The reason being, that the likelihood (equation \ref{eq::gn},
specifically $(F(x) - q)^2$) does not change much for a large change
in $x$. This is the case, in particular, in the tails. Therefore,
a bad estimate, e.g., of $q = F(x) = 0.0001$ for $q_{\text{true}} = 0.001$ 
leads to almost the same likelihood score as the correct value
$q = F(x) = 0.001 = q_{\text{true}}$.
Because of that, the CDF regression with Gaussian noise is
totally not suitable for applications where tails of the
distribution are important. Order statistics is a better 
choice here.

\subsection{Real World Data}

\begin{table}[t]
    \caption{Aggregated salary data of some European countries from 2016.
            Columns 25, 50 and 75 are the corresponding quantiles of the
            yearly salary of individuals. ``Sample Size'' indicates
            number of total samples of the survey.}
    \label{salaryData}
    \vskip 0.15in
    \begin{center}
    \begin{small}
    \begin{sc}
    \begin{tabular}{ccccc}
        \toprule
        Country & Sample Size & 25 & 50 & 75 \vspace{0.1cm} \\
        \midrule
        EL &  12918 &   4930 &   7500 &  11000 \\
        ES &  19177 &   8803 &  13681 &  20413 \\
        FR &  21325 &  16185 &  21713 &  29008 \\
        IT &  24969 &  10699 &  16247 &  22944 \\
        LU &  10292 &  23964 &  33818 &  48692 \\
        NL &  12748 &  16879 &  22733 &  30327 \\
        SE &  11635 &  17794 &  25164 &  33365 \\
        UK &  17645 &  14897 &  21136 &  30151 \\
        \bottomrule
    \end{tabular}
    \end{sc}
    \end{small}
    \end{center}
    \vskip -0.1in
\end{table}

\begin{table*}
    \caption{The mean of the log likelihood samples for different 
            models and countries. Best fits to the salary quantile 
            data (bold numbers) are given by the Weibull, lognormal 
            and gamma distribution. The + and - values show the
            distance to the 95 and 5 \% quantile of the posterior
            log-likelihood distribution.}
    \label{salaryFit}
    \vskip 0.15in
    \begin{center}
    \begin{small}
    \begin{sc}
    \begin{tabular}{crrrrrrr}
        \toprule
        country  &                    weibull &                  lognormal &                     gamma &                   inv\_gamma &                     frechet &                   chi\_square &                  exponential \\
        \midrule
        EL &   $-6.9^{\ +0.9}_{\ -2.1}$ &    $4.3^{\ +0.9}_{\ -2.0}$ &  ${\bf 10.2}^{\ +1.0}_{\ -2.0}$ &   $-31.5^{\ +0.9}_{\ -2.0}$ &   $-81.1^{\ +1.0}_{\ -2.2}$ &  $-2063.9^{\ +0.5}_{\ -1.3}$ &  $-1416.4^{\ +0.5}_{\ -1.3}$ \vspace{0.1cm}\\
        ES &  $-13.4^{\ +1.0}_{\ -1.9}$ &   $-0.2^{\ +1.0}_{\ -2.1}$ &  ${\bf 10.1}^{\ +0.9}_{\ -1.8}$ &   $-58.9^{\ +1.0}_{\ -2.0}$ &  $-130.4^{\ +1.0}_{\ -2.1}$ &  $-2776.2^{\ +0.5}_{\ -1.4}$ &  $-1854.7^{\ +0.5}_{\ -1.4}$ \vspace{0.1cm}\\
        FR &  $-57.7^{\ +1.0}_{\ -2.0}$ &   ${\bf 13.0}^{\ +0.9}_{\ -2.0}$ &   $3.5^{\ +1.0}_{\ -2.2}$ &    $-4.4^{\ +1.0}_{\ -2.1}$ &   $-76.8^{\ +0.9}_{\ -2.0}$ &  $-5847.6^{\ +0.5}_{\ -1.4}$ &  $-4554.3^{\ +0.5}_{\ -1.4}$ \vspace{0.1cm}\\
        IT &    ${\bf 9.1}^{\ +0.9}_{\ -2.0}$ &  $-48.9^{\ +0.9}_{\ -2.0}$ &   $5.3^{\ +0.9}_{\ -2.0}$ &  $-155.2^{\ +0.9}_{\ -2.0}$ &  $-290.0^{\ +1.0}_{\ -2.1}$ &  $-4500.3^{\ +0.5}_{\ -1.4}$ &  $-3139.8^{\ +0.5}_{\ -1.4}$ \vspace{0.1cm}\\
        LU &  $-47.7^{\ +1.0}_{\ -2.0}$ &    ${\bf 9.3}^{\ +1.0}_{\ -2.1}$ &  $-9.3^{\ +0.9}_{\ -2.1}$ &     $9.1^{\ +0.9}_{\ -2.0}$ &   $-10.9^{\ +0.9}_{\ -1.9}$ &  $-2062.3^{\ +0.5}_{\ -1.4}$ &  $-1524.1^{\ +0.5}_{\ -1.4}$ \vspace{0.1cm}\\
        NL &  $-23.3^{\ +0.9}_{\ -1.9}$ &   ${\bf 11.6}^{\ +1.0}_{\ -2.0}$ &   $9.0^{\ +0.9}_{\ -1.9}$ &    $-1.9^{\ +1.0}_{\ -2.0}$ &   $-49.4^{\ +1.0}_{\ -2.1}$ &  $-3473.8^{\ +0.5}_{\ -1.4}$ &  $-2698.1^{\ +0.5}_{\ -1.4}$ \vspace{0.1cm}\\
        SE &   ${\bf 11.4}^{\ +1.0}_{\ -2.0}$ &  $-21.2^{\ +0.9}_{\ -2.0}$ &   $3.9^{\ +1.0}_{\ -2.0}$ &   $-63.0^{\ +1.0}_{\ -2.0}$ &  $-138.7^{\ +0.9}_{\ -2.0}$ &  $-2910.8^{\ +0.5}_{\ -1.3}$ &  $-2191.0^{\ +0.5}_{\ -1.4}$ \vspace{0.1cm}\\
        UK &  $-62.8^{\ +0.9}_{\ -2.0}$ &   ${\bf 12.1}^{\ +1.0}_{\ -2.0}$ &  $-8.1^{\ +1.0}_{\ -2.0}$ &     $0.5^{\ +0.9}_{\ -1.9}$ &   $-45.6^{\ +0.9}_{\ -1.9}$ &  $-3582.7^{\ +0.5}_{\ -1.3}$ &  $-2641.1^{\ +0.5}_{\ -1.3}$ \vspace{0.1cm}\\
        \bottomrule
    \end{tabular}
    \end{sc}
    \end{small}
    \end{center}
    \vskip -0.1in
\end{table*}

Due to data privacy acts, many times the data are not available on an
individual basis but only in aggregated forms. Whenever this is the case,
traditional machine learning techniques are not very useful.
In the case of medical data, one cannot simply publish private
data of individual patients. But the the data are published in aggregated form.
The same is true for other sensitive data sources, e.g.: economic data.
Here also data is provided in aggregated form.
As an example we take the distribution of income of European countries.
Table \ref{salaryData} shows the salary values for 
25, 50 and 75\% quantiles and the total number of samples of 
the survey\footnote{The data are downloaded from the Eurostat homepage 
    (Distribution of income by quantiles - EU-SILC and ECHP surveys)
    at \href{https://ec.europa.eu/eurostat/}{https://ec.europa.eu/eurostat/}.
    Information about the sample size is also available on the website
    (EU and national quality reports).
}.

\begin{figure}
 \centering
 \includegraphics[width=0.99\textwidth]{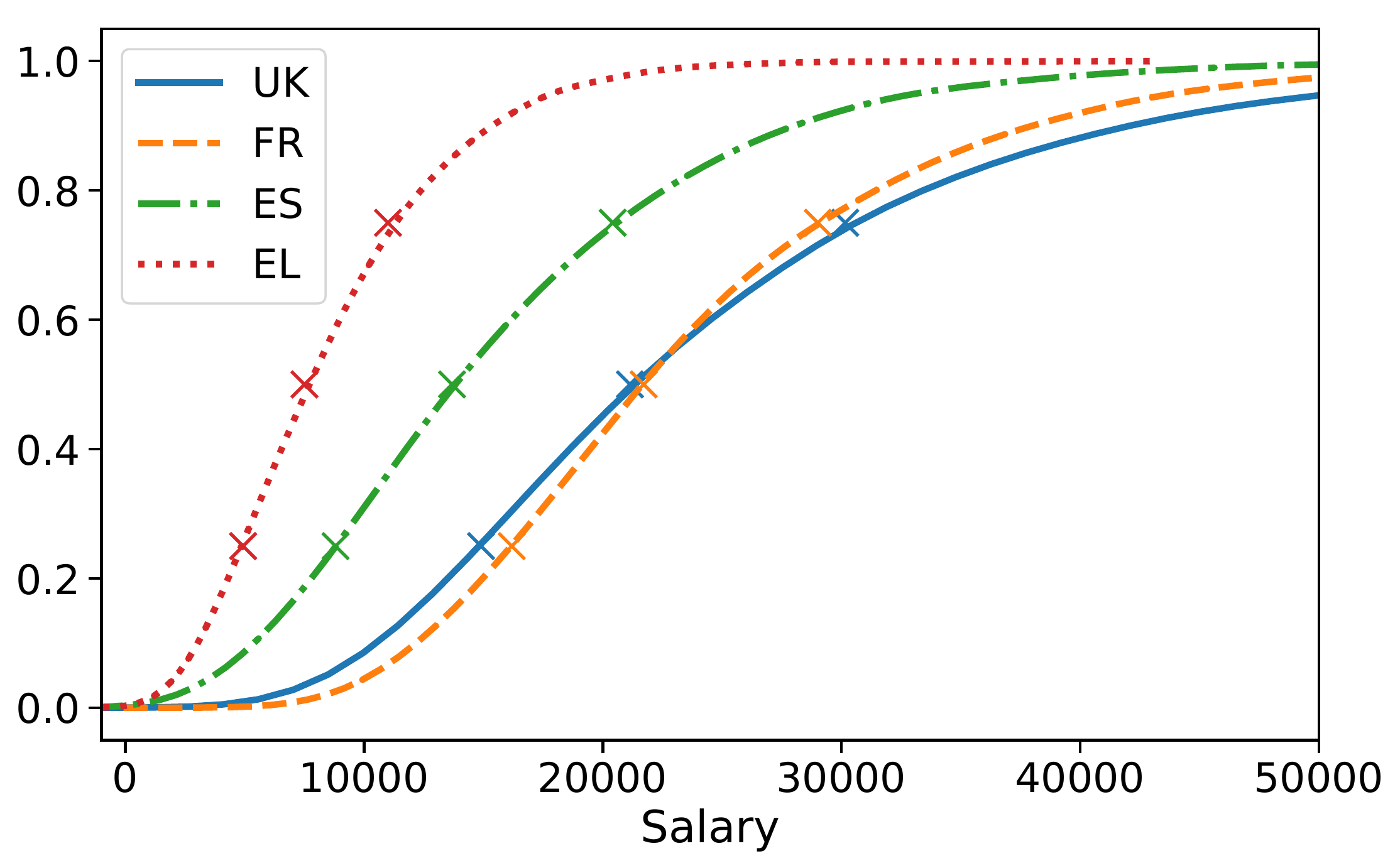}
 \caption{Posterior predictive cumulative distribution.
      Observed data are marked by crosses. 
      Different colors indicate observed data and learned distribution for 
      different countries.}
 \label{predictive_dist}
\end{figure}

Before fitting the data as explained in section \ref{OrderStatistics}, we 
normalize it by dividing the salaries by the median (50\% quantile) of 
each country. The normalized data are then fitted by various distributions.
The goodness of the fits measured by the log-likelihood
is shown in table \ref{salaryFit}. Each row corresponds to 
one particular country and contains the log-likelihood values for
several models. The highest log-likelihood value (best fit to the
data from the particular country) is emphasised 
in bold writing. The corresponding fit for the cumulative density 
(posterior predictive cumulative density) of the salary
is also shown in Figure \ref{predictive_dist}, i.e.
\begin{equation}
    P(X < x') = \int F_{\Bv{\theta}}(x') p(\Bv{\theta}|\Bv{x} \; ,
    \Bv{q}, N) \text{d}\Bv{\theta} \; .
\end{equation}

\citet{Bandourian2002} suggests to use a Weibull distribution to 
fit salary data. However, the data they use are not as aggregated as ours are.
Table \ref{salaryFit} shows that Weibull as well as lognormal and gamma 
distributions provide reasonable fits to the data. 
However, also in this case we should keep
in mind that we try to fit the whole distribution from 
just 3 quantiles, which might not contain enough information 
to pinpoint one single distribution.
Given the predictive distribution, questions like the following
can be answered very easily: What is the threshold for the 99\%
quantile (earnings of the top 1\%)?
In the supplementary material we provide a table estimating the
99\% quantile of the predictive distribution
for the best fits for each country.
In addition, a similar plot to Figure \ref{xt_xpred_misspecified} is
included in the supplementary material for UK. While being based on
three quantiles $q = 0.25, 0.5, 0.75$ only (where we know the empirical salaries)
the log-normal model shows no sign of misspecification in contrast to
the other considered models.

\section{Conclusion}
\label{Conclusion}
We proposed an alternative approach for quantile matching
estimation. In contrast to the widely used procedure of minimzing the
MSE between the theoretical and empirical CDF, our Ansatz is based on
the order statistics of the samples. This leads to a principled noise
model combining the information contributed from several quantiles.
In particular, we showed that our model correctly accounts for the
higher uncertainty of tail quantiles, whereas the Gaussian noise model
-- corresponding to MSE minimization -- overemphasises the central
part of the distribution. Furthermore, our Bayesian approach allows
for a principled assessment of model uncertainty and model comparison.
As an example, we fitted income data from several European countries
with our proposed method and compared several candidate
distributions.

\bibliography{order}
\bibliographystyle{apalike}

\clearpage

\appendix
\section*{\LARGE{Supplementary Material}}

\section{Normalization Constant}
\label{ANormalizationConstatnt}
To verify that equation (\ref{normalizationConstant}) is indeed the
normalization constant, we have to integrate equation 
(\ref{jointUniformOS}) with respect to all $u$'s
\begin{equation}
    \frac{1}{c} = \int_{0}^{u_2} \int_{u_1}^{u_3} \dots \int_{u_{M-1}}^{1}
    u_1^{k_1-1} (1-u_M)^{n-k_M}
    \prod_{m=2}^M (u_m - u_{m-1})^{k_m-k_{m-1}-1}
    \text{d}u_1 \text{d}u_2 \dots \text{d}u_M \; .
\end{equation}
For the integration of a particular $u_i$, however, only the following term is
relevant
\begin{equation}
    \int_{u_{i-1}}^{u_{i+1}} (u_i - u_{i-1})^{k_i-k_{i-1}-1}
    (u_{i+1} - u_i)^{k_{i+1}-k_i-1} \text{d} u_i \; .
\end{equation}
The rest is constant with respect to $u_i$. To solve this integral
we substitute $u = \frac{u_i - u_{i-1}}{u_{i+1} - u_{i-1}}$ for $u_i$
and get
\begin{align}
    \int_{u_{i-1}}^{u_{i+1}} (u_i &- u_{i-1})^{k_i-k_{i-1}-1}
    (u_{i+1} - u_i)^{k_{i+1}-k_i-1} \text{d} u_i \nonumber \\
    &= (u_{i+1} - u_{i-1})^{k_{i+1} - k_{i-1} - 1 }
    \int_0^1 u^{k_i-k_{i-1}-1} (1-u)^{k_{i+1} - k_i-1} du
    \nonumber \\
    &= (u_{i+1} - u_{i-1})^{k_{i+1} - k_{i-1} - 1 }
    \frac{\Gamma(k_i-k_{i-1})\Gamma(k_{i+1}-k_{i})}
    {\Gamma(k_{i+1}-k_{i-1})} \; .
\end{align}
Note that the integrand in the second step is a unnormalized beta
distribution, where we already know the normalization constant
\citep{bishop2006}.
$\Gamma(\cdot)$ is the Gamma-function, which, for integer input 
has the form $\Gamma(n+1) = n!$.

By integrating out $u_i$, the resulting expression has still 
a similar form.
Thus, by successive applications of the above result, we obtain
the normalization constant as in equation (\ref{normalizationConstant}).

\clearpage
\section{Stan code for Bayesian Quantile Matching Estimation}

Below is the stan code to fit a Weibull distribution to the data.

\begin{lstlisting}[style=custom]
  functions{
    real orderstatistics(int N, int M, vector q, vector U){
      real lpdf = 0;
      lpdf += lgamma(N+1) - lgamma(N*q[1]) - lgamma(N-N*q[M]+1);
      lpdf += (N*q[1]-1)*log(U[1]);
      lpdf += (N-N*q[M])*log(1-U[M]);
      for (m in 2:M){
        lpdf += -lgamma(N*q[m]-N*q[m-1]);
        lpdf += (N*q[m]-N*q[m-1]-1)*log(U[m]-U[m-1]);
      }
      return lpdf;
    }
  }
  data{
    int N;          // total sample size
    int M;          // number of observed quantiles
    vector[M] q;    // quantiles
    vector[M] X;    // quantile values
  }
  parameters{
    real<lower=0> shape;
    real<lower=0> scale;
  }
  transformed parameters{
    vector[M] U;
    for (m in 1:M)
      U[m] = weibull_cdf(X[m], shape, scale);
  }
  model{
    shape ~ normal(0, 100);
    scale ~ normal(0, 100);
    target += orderstatistics(N, M, q, U);
    for (m in 1:M)
      target += weibull_lpdf(X[m] | shape, scale);
  }
  generated quantities {
    real<lower=0> predictive_dist = weibull_rng(shape, scale);
    real log_prob = orderstatistics(N, M, q, U);
    for (m in 1:M)
      log_prob += weibull_lpdf(X[m] | shape, scale);
  }
\end{lstlisting}

To fit a lognormal, for example, instead of a weibull, we only need to
change the following lines: \\
Line 27 to 
\begin{lstlisting}[style=oneline] 
  U[m] = lognormal_cdf(X[m], mu, sigma);
\end{lstlisting}
Line 34 to
\begin{lstlisting}[style=oneline] 
  target += lognormal_lpdf(X[m] | mu, sigma);
\end{lstlisting}
Line 37 to
\begin{lstlisting}[style=oneline] 
  real<lower=0> predictive_dist = lognormal_rng(mu, sigma);
\end{lstlisting}
Line 40 to
\begin{lstlisting}[style=oneline] 
  log_prob += lognormal_lpdf(X[m] | mu, sigma);
\end{lstlisting}
In addition, we assume that the \texttt{shape} and \texttt{scale} parameters have been renamed to 
\texttt{mu} and \texttt{sigma} within the code.

\clearpage

\section{Predicted earnings of the top 1\% of the population}

\begin{table}[ht!]
    \caption{The earnings of the top 1\% according to the best
    model based on the log-likelihood score. The + and - values show the
    distance to the 95 and 5 \% quantile of the posterior
    predictive distribution.}
    \label{top1p}
    \vskip 0.15in
    \begin{center}
    \begin{small}
    \begin{sc}
    \begin{tabular}{ccr}
        \toprule
        Country & best model &                        99\% quantile \\
        \midrule
        EL &      gamma &     $23268.6^{\ +406.8}_{\ -400.4}$ \vspace{0.1cm}\\
        ES &      gamma &     $44343.5^{\ +701.7}_{\ -633.7}$ \vspace{0.1cm}\\
        FR &  lognormal &     $59331.9^{\ +834.6}_{\ -838.8}$ \vspace{0.1cm}\\
        IT &    weibull &     $41096.2^{\ +483.8}_{\ -467.6}$ \vspace{0.1cm}\\
        LU &  lognormal &  $115693.5^{\ +3038.5}_{\ -2796.1}$ \vspace{0.1cm}\\
        NL &  lognormal &   $62265.1^{\ +1185.3}_{\ -1142.6}$ \vspace{0.1cm}\\
        SE &    weibull &     $53926.5^{\ +754.0}_{\ -744.5}$ \vspace{0.1cm}\\
        UK &  lognormal &   $71466.4^{\ +1294.2}_{\ -1280.7}$ \vspace{0.1cm}\\
        \bottomrule
    \end{tabular}
    \end{sc}
    \end{small}
    \end{center}
    \vskip -0.1in
\end{table}

\section{In sample and out of sample predictions for UK}

\begin{figure}[ht!]
 \centering
 \includegraphics[width=0.90\textwidth]{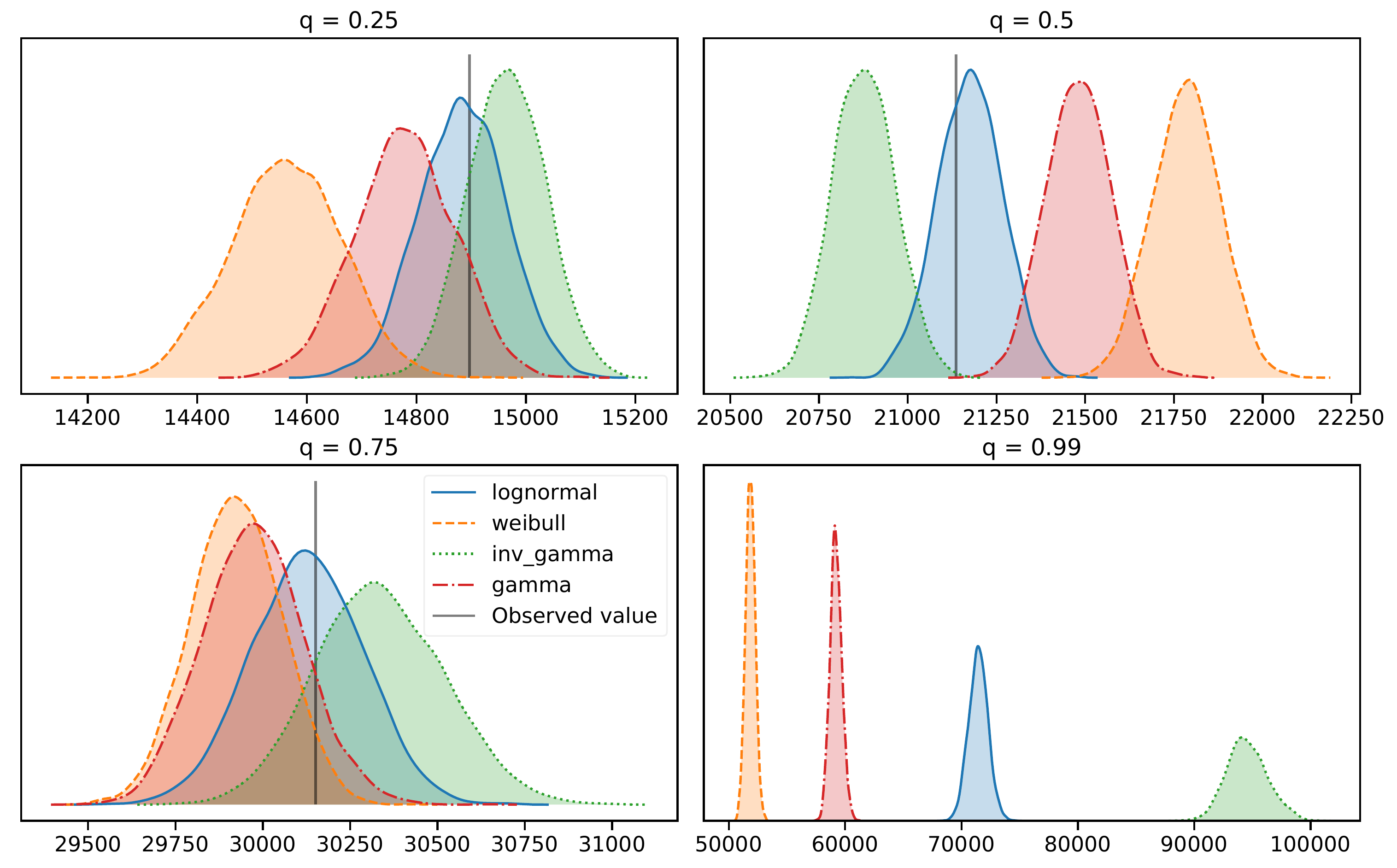}
 \caption{The top row and the left side of the bottom row shows the
 predictions made for the quantile that we also observed and the
 bottom right plot shows an out of sample prediction for the 99\% quantile.
 Different colors indicate different models. The best model according to
 the log-likelihood value was the lognormal distribution. This figure
 visually validates that result.}
 \label{xx}
\end{figure}

\end{document}